\documentclass{PoS}
\newcommand{\Ef}{\mathcal{E}}

\newcommand{\EA}{\text{EN1}}
\newcommand{\EB}{\text{EN2}}
\newcommand{\EC}{\text{EN3}}
\newcommand{\ED}{\text{EN4}}

\newcommand{\EEA}{\text{EN5}}
\newcommand{\EEB}{\text{EN6}}
\newcommand{\EEC}{\text{EN7}}
\newcommand{\EED}{\text{EN8}}

\usepackage{amsmath}

\usepackage[normalem]{ulem}
\usepackage{color}

\title{Finite volume study of electric polarizabilities from lattice QCD.}

\ShortTitle{Finite volume study of electric polarizabilities from lattice QCD.}

\author{\speaker{Michael Lujan},~Andrei Alexandru, Walter Freeman,~and~Frank Lee\\
        The George Washington University, Washington DC, USA\\
        E-mail: \email{mlujan@gwmail.gwu.edu},~~ \email{aalexan@gwu.edu},~~ \email{wfreeman@gwu.edu},~~ \email{fxlee@gwu.edu}}

\abstract{
Knowledge of the electric polarizability is crucial to understanding the interactions of hadrons with electromagnetic fields. The neutron polarizability is very sensitive to the quark mass and is expected to diverge in the chiral limit. Here we present results for the electric polarizability of the neutron, neutral pion, and neutral kaon on eight ensembles with nHYP-smeared clover dynamical fermions with two different pion masses (227 and 306 MeV). These are currently the lightest pion masses used in polarizability studies.  For each pion mass we compute the polarizability at four different volumes and perform an infinite volume extrapolation for the three hadrons. Along with the infinite volume extrapolation we conduct a chiral extrapolation for the kaon polarizability to the physical point. We compare our results for the neutron polarizability to predictions from chiral perturbation theory.
} 

\FullConference{32nd International Symposium on Lattice Field Theory - LATTICE 2014\\
		June 23 - June 28, 2014\\
	         New York, USA}

\begin{document}

\section{Introduction}

To lowest order, the response of a composite particle to an electromagnetic field can be parameterized by the effective Hamiltonian:
\begin{equation}
\mathcal{H}_{em} = -\vec{p}\cdot\vec{\Ef} -\vec{\mu}\cdot\vec{B} -\frac{1}{2}\left(\alpha \Ef^2 + \beta B^2\right)+...,
\end{equation}
where $\vec{p}$ and $\vec{\mu}$ are the electric and magnetic dipole moments, respectively, and $\alpha$ and $\beta$ are the electric and magnetic polarizabilities.  Due to time-reversal symmetry of the strong interaction the electric dipole moment vanishes. Furthermore, by considering the simplified case of a weak constant electric field, the leading order interaction comes from the  polarizability term at $\mathcal{O}(\Ef^2)$. The polarizability is a first-order structure constant which measures the rigidity of the hadron in the presence of the external field.  

%An accurate determination of the polarizability can lead to better constraints on other interesting nuclear physics phenomena. For example, the leading uncertainty in the proton radius puzzle is due to the relatively large uncertainties in the proton polarizability~\cite{Pohl:2010fk}. Another example is the neutron-proton mass splitting calculation~\cite{WalkerLoud:2012bg}. In this case, the isovector-nucleon magnetic polarizability accounts for the largest error in the calculation.

In this work, we use lattice QCD to compute the electric polarizability ($\alpha$). We employ the background field method to extract the polarizability. Previous lattice calculations~\cite{Engelhardt:2007ub,Detmold:2010ts,Detmold:2009dx,Fiebig:1988en,Alexandru:2009id,Alexandru:2008sj} were done at relatively heavy pion masses leaving the chiral region largely unexplored. Here, we use 2-flavor n-HYP clover fermions with two different pion masses (227 MeV and 306 MeV) to study the chiral behavior of the polarizability. Moreover, for each mass we compute $\alpha$ on four different lattice volumes to study the volume dependence.  

We analyze three neutral hadrons: neutral pion, neutral kaon, and the neutron. For each hadron we performed an infinite volume extrapolation.  For the kaon, we also performed a chiral extrapolation to the physical point.  The results of our neutron polarizability, will be compared to predictions from chiral perturbation theory.   We note that our work, though done on dynamical configurations, uses electrically-neutral sea quarks throughout. 

The outline of the paper is as follows: In section~\ref{sec:methodology} we discuss the lattice methodology to extract the polarizability and our fitting procedure. In section~\ref{sec:results} we present our results for the neutron, neutral pion and neutral kaon polarizability. We then summarize and present an outlook in section~\ref{sec:conclusion}.

\section{Methodology}\label{sec:methodology}

\subsection{Background field method}
We use the background field method to place the electric field on the lattice. The method uses minimal coupling which augments the static electromagnetic vector potential ($A_{\mu}$) to the covariant derivative, i.e.
\begin{equation}
D_{\mu} = \partial_{\mu} -igG_{\mu} -iqA_{\mu},
\end{equation}
where $G_{\mu}$ are the gluon field degrees of freedom. In practice, this amounts to an overall multiplicative phase factor to the original gauge links which appear in $D_{\mu}$:
\begin{equation}
U_{\mu} \rightarrow e^{-iqaA_{\mu}} U_{\mu}.
\end{equation}
%%
%%
%To achieve a constant electric field, in say the $x$-direction, we may choose $A_{x} = -i\Ef t$ which provides us with a real multiplicative factor. The additional factor of $i$ comes from the fact that we use a Euclidean metric on the lattice. A more convenient choice, and the one implemented in this study,  is the use of an imaginary value for the electric field which leads to a U(1) multiplicative factor that keeps the links unitary. When using an imaginary value of the field, the energy shift due to the polarizability acquires an additional negative sign leading to a positive energy shift for a positive value of the polarizability \cite{Alexandru:2008sj}:
%%
%\begin{equation}
%\delta E = -\frac{1}{2} \alpha \Ef^2~ \rightarrow ~ \delta E = +\frac{1}{2} \alpha \Ef^2
%\end{equation}
%%%
The polarizability is extracted by computing the variation of the hadron's ground state energy with and without the presence of an electric field. 
 
In order to extract the polarizability we need to use a weak enough electric field so that higher order terms in the field 
expansion can be safely neglected. 
%\andrei{It turns out that if we use periodic boundary conditions, for our lattice sizes, 
%we need to use relatively large electric fields}{}. 
In this work we instead use Dirichlet boundary conditions (DBC). 
The advantage of this is that we can use arbitrarily small values of the field. DBC also protects against the 
%\andrei{so-called}{vacuum instability due to the} 
vacuum instability due to the Schwinger 
mechanism~\cite{Schwinger:1951nm}.\footnote{This instability only
occurs for real electric fields and not for imaginary electric fields as used in most lattice studies. However,
to use imaginary fields we need to rely on the analyticity of the theory around the point where the electric field is
zero. Schwinger mechanism signals that this is not an analytical point generically. DBC offers one way to restore this
analyticity.}
However, using DBC creates boundary effects. One of them is the introduction of a non-zero momentum for 
%\andrei{which alters}{for} 
the hadron of magnitude
%\andrei{'s energy by a factor of}{ of magnitude} 
$\pi/L$.  The lowest energy state of the system is then $E = \sqrt{m^2 + (\pi/L)^2}$, where $m$ is the mass of the hadron. 
%\andrei{Thus, when we compute the variation of the hadron's energy there is an unwanted portion associated with the non-zero momentum.}{} 
To account for this motion 
we compute the mass shift ($\delta m$) motion due to the polarizability 
%\andrei{}{motion} 
%\andrei{}{due to polarizability} 
using the relation
\begin{equation}
\delta m = \delta E \frac{E}{m}.
\end{equation}
The hadron's mass ($m$) is computed using periodic boundary conditions.

\subsection{Fitting Procedure} \label{sec:fitprocedure}

%As already mentioned, we are computing the polarizability for neutral hadrons (charged hadrons have additional complications associated with the fact that they accelerate in an electric field). 
The form of the correlators, for neutral hadrons in an constant electric field, retain their exponential fall off, 
allowing us to use some of the standard spectroscopy techniques to measure the shift in hadrons' energies.  

The main difference in the fitting analysis is the fact that we need to extract the energy shift from the three correlation functions: $G_0, G_{+\eta}$, and $G_{-\eta}$, which are the correlation functions for the zero-field, and non-zero fields in the positive and negative $x$-direction, respectively. Since all three correlators are computed from the set of gauge configurations they are highly correlated; we therefore need to properly account for the correlations among them. To do this we construct the following difference vector as
\begin{eqnarray}
\mathbf{v}_{i} &\equiv& f(t_i) - \langle G_{0}(t_i)\rangle,\\
\mathbf{v}_{N+i} &\equiv&  \bar{f}(t_{i}) - \langle G_{+\Ef}(t_{i})\rangle, \nonumber\\
\mathbf{v}_{2N+i} &\equiv&  \bar{f}(t_{i}) - \langle G_{-\Ef}(t_{i})\rangle~~~\mbox{for}~i=1,...,N \nonumber
\end{eqnarray}
where  $t_1...t_N$ is the fit window, $f(t) = A~e^{-E t}$ and $\bar{f}(t) =(A+\delta A)~e^{-(E+\delta E)t}$. We minimize the $\chi^2$ function, 
\begin{equation}
\chi^2= \mathbf{v}^{\mathbf{\dagger}}~ \mathbf{C}^{-1}~ \mathbf{v}, \nonumber
\end{equation}
in the usual fashion, where $\mathbf{C}$ is the $3N~\times~3N$ correlation matrix which encodes the correlations among the three different correlators.

%has the block structure
%\[ \mathbf{C} = \left( \begin{array}{ccc}
%C_{0 0} ~& C_{0 +}~ & C_{0 -} \\
%C_{+ 0} ~& C_{+ +}~ & C_{+ -} \\
%C_{- 0} ~& C_{- +} ~& C_{- -} \end{array} \right),\] 
%where $0,+,-$ represent  $G_0, G_{+\Ef}$, and $G_{-\Ef}$ respectively. Note that the symmetrization is done implicitly in this procedure, since $\bar{f}$ is the same for  $G_{+\Ef}$, and $G_{-\Ef}$. This method is used to extract all parameters presented in this work.

\section{Ensemble Details and Results}\label{sec:results}

%%%
\begin{table}[t]
\begin{center}
\begin{tabular}{c c c c c c}
\hline
Ensemble &Lattice&  $a$ (fm) &  $\kappa$& $N_{\text{c}}$&\\ 
\hline
\hline
$\EA$&$16\times16^2 \times 32$  & 0.1245& 0.12820  &  230\\
$\EB$&$24\times 24^2 \times 48$ & 0.1245& 0.12820  &  300\\
$\EC$&$30 \times 24^2 \times 48$& 0.1245& 0.12820 &  300\\
$\ED$&$48 \times 24^2 \times 48$& 0.1245& 0.12820 &  270\\
\hline
\hline
$\EEA$&$16\times16^2 \times 32$& 0.1215& 0.12838  & 230\\
$\EEB$&$24\times 24^2 \times 64$& 0.1215& 0.12838 & 450\\
$\EEC$&$28\times 24^2 \times 64$& 0.1215& 0.12838 & 670\\
$\EED$&$32\times 24^2 \times 64$& 0.1215& 0.12838 & 500\\
\hline
\end{tabular}
\end{center}
\caption{Details of the lattice ensembles used in this work. $N_c$ is the number of configurations. The separation between the top four ensembles and bottom four ensembles is used to indicate the two different sea quark masses more clearly. }
\label{tab:ensemblesres}
\end{table}
%%%

In this work we use 2-flavor nHYP-clover fermions~\cite{Hasenfratz:2007rf} with two different pion masses (227 MeV and 306 MeV) in order to study the chiral behavior of the polarizability. For each pion mass we compute $\alpha$ on four different volumes in order to study the volume effects of the polarizability. A description of the ensembles are tabulated in Table~\ref{tab:ensemblesres}. In our simulations we use a field size of $\eta \equiv a^2 q_d \Ef = 10^{-4}$, where $q_d$ is the charge of the down quark. For a detailed discussion for the choice of $\eta$ we refer the reader to 
our previous work~\cite{Lujan:2013qua}. 

\subsection{Finite Volume Effects}

%%%%%%%%%
 \begin{figure}[t]
 \centering
 \includegraphics[width= 2.6 in]{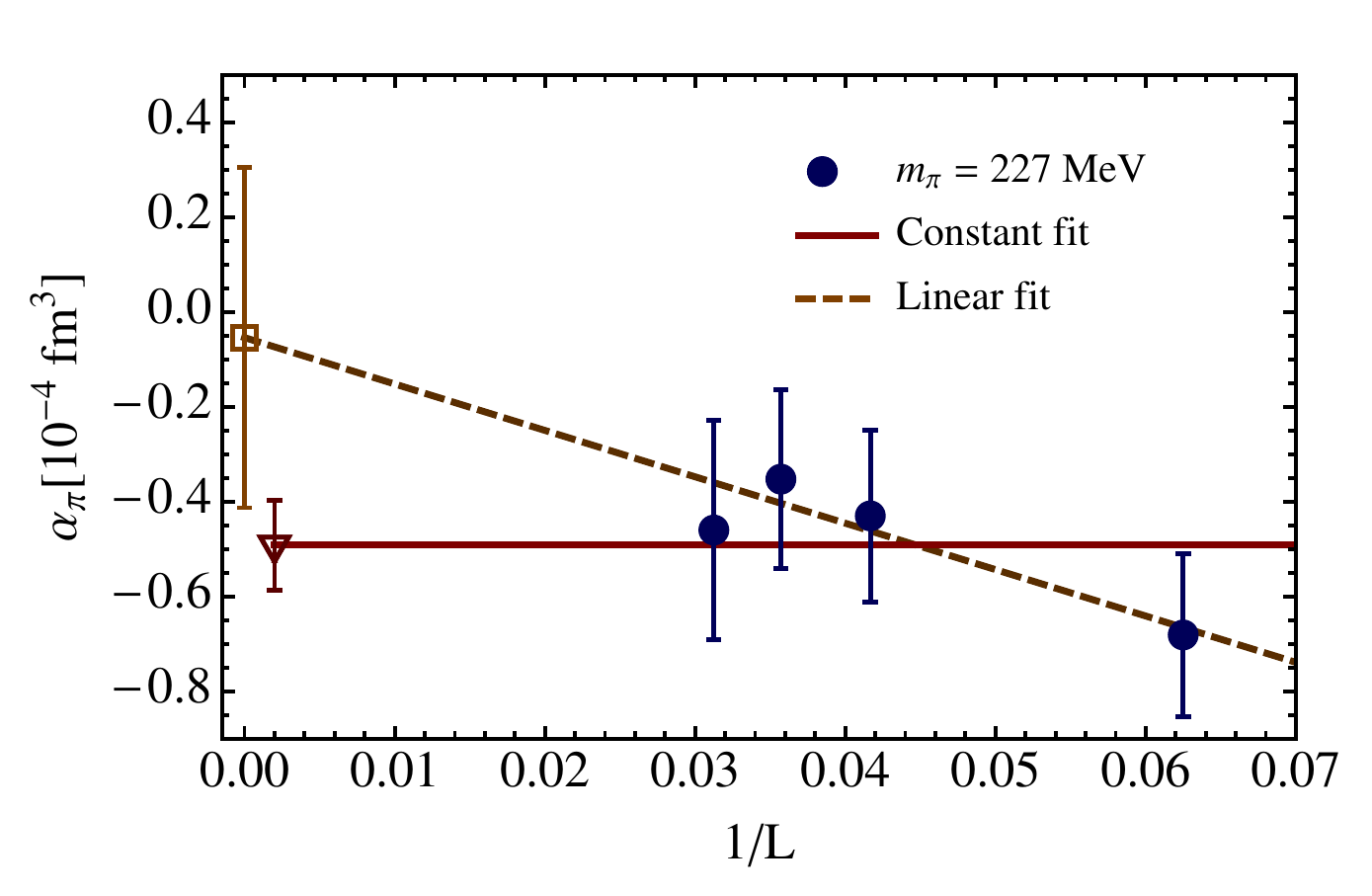}
  \includegraphics[width= 2.6 in]{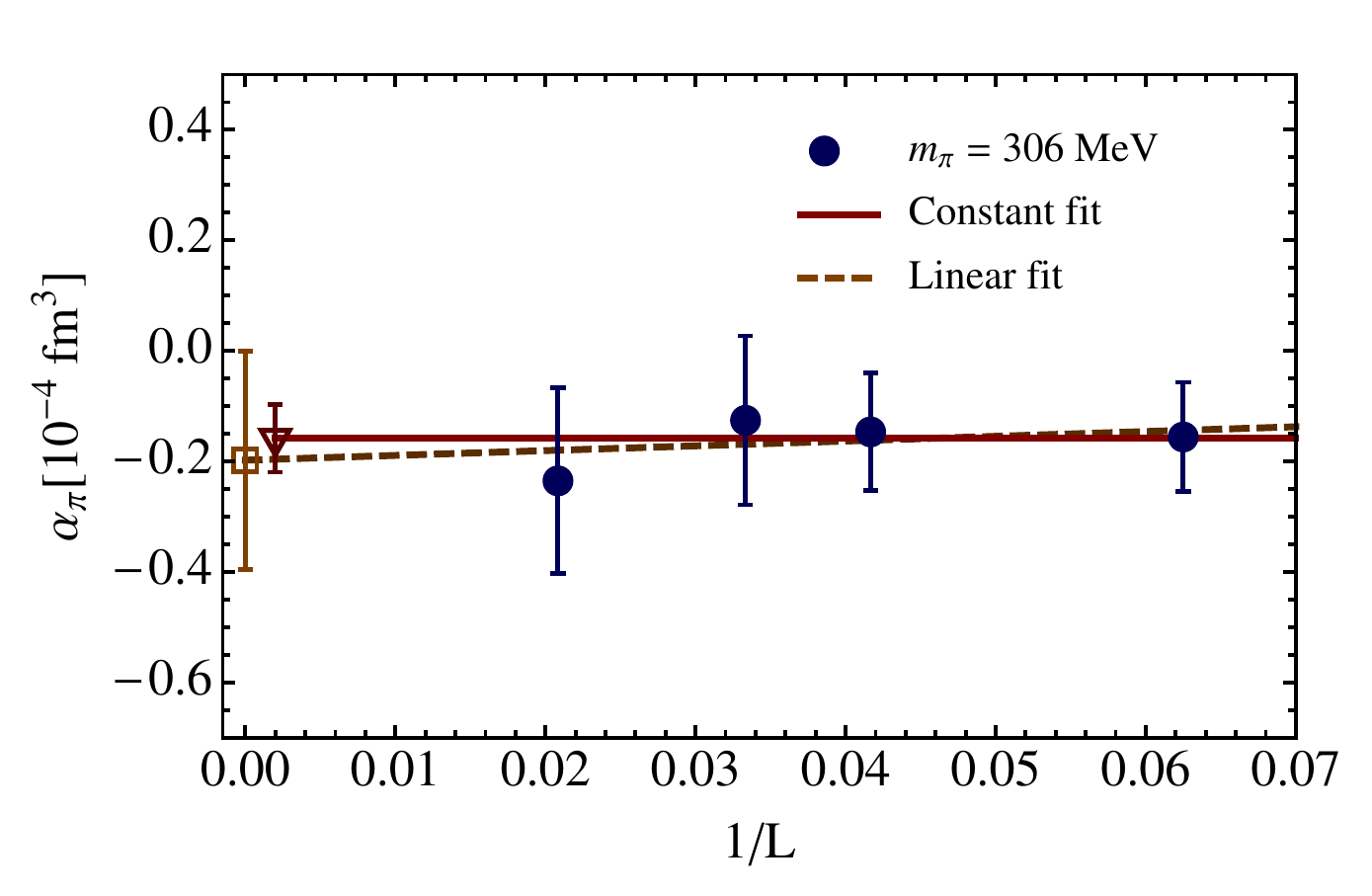}
   \includegraphics[width= 2.6 in]{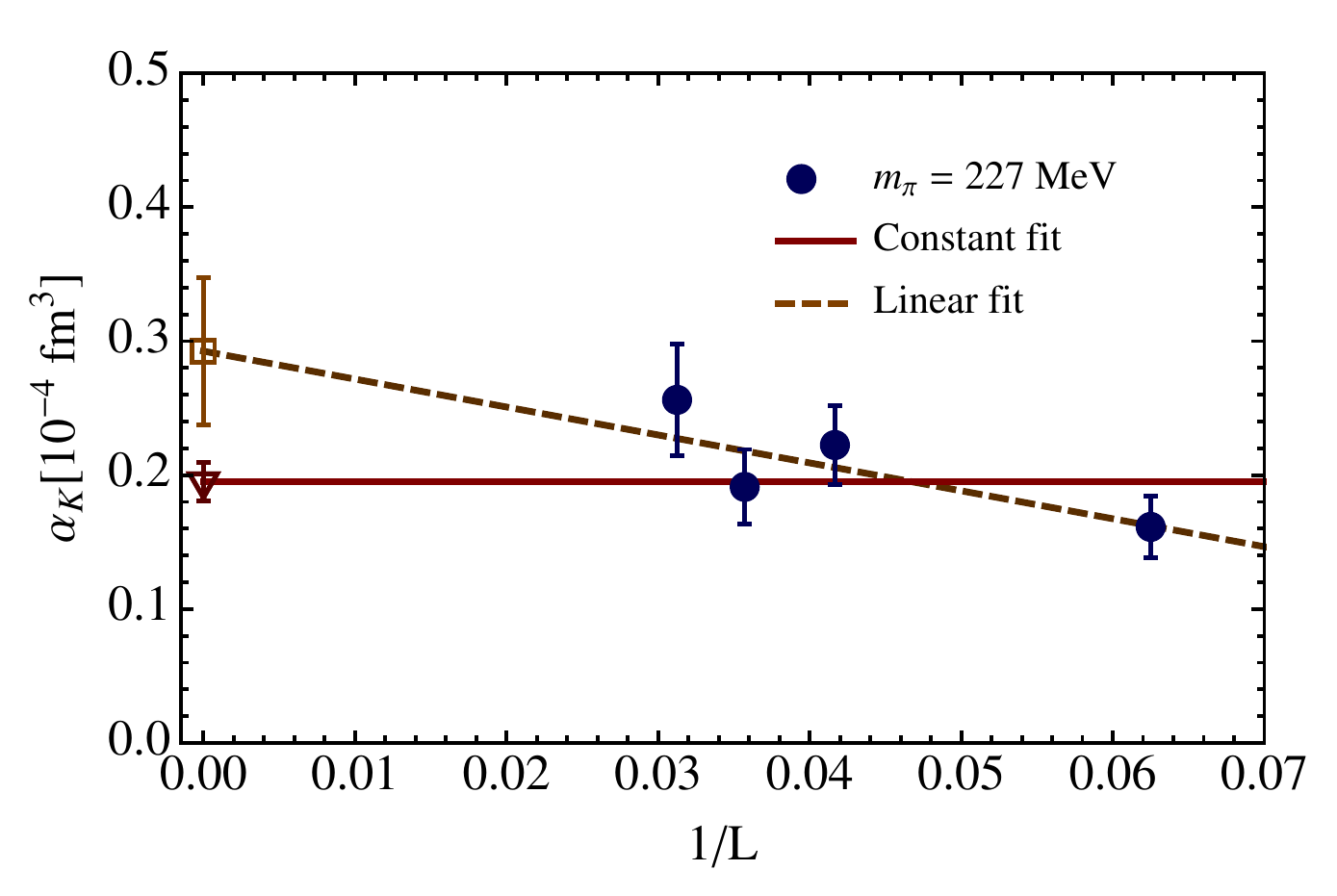}
  \includegraphics[width= 2.6 in]{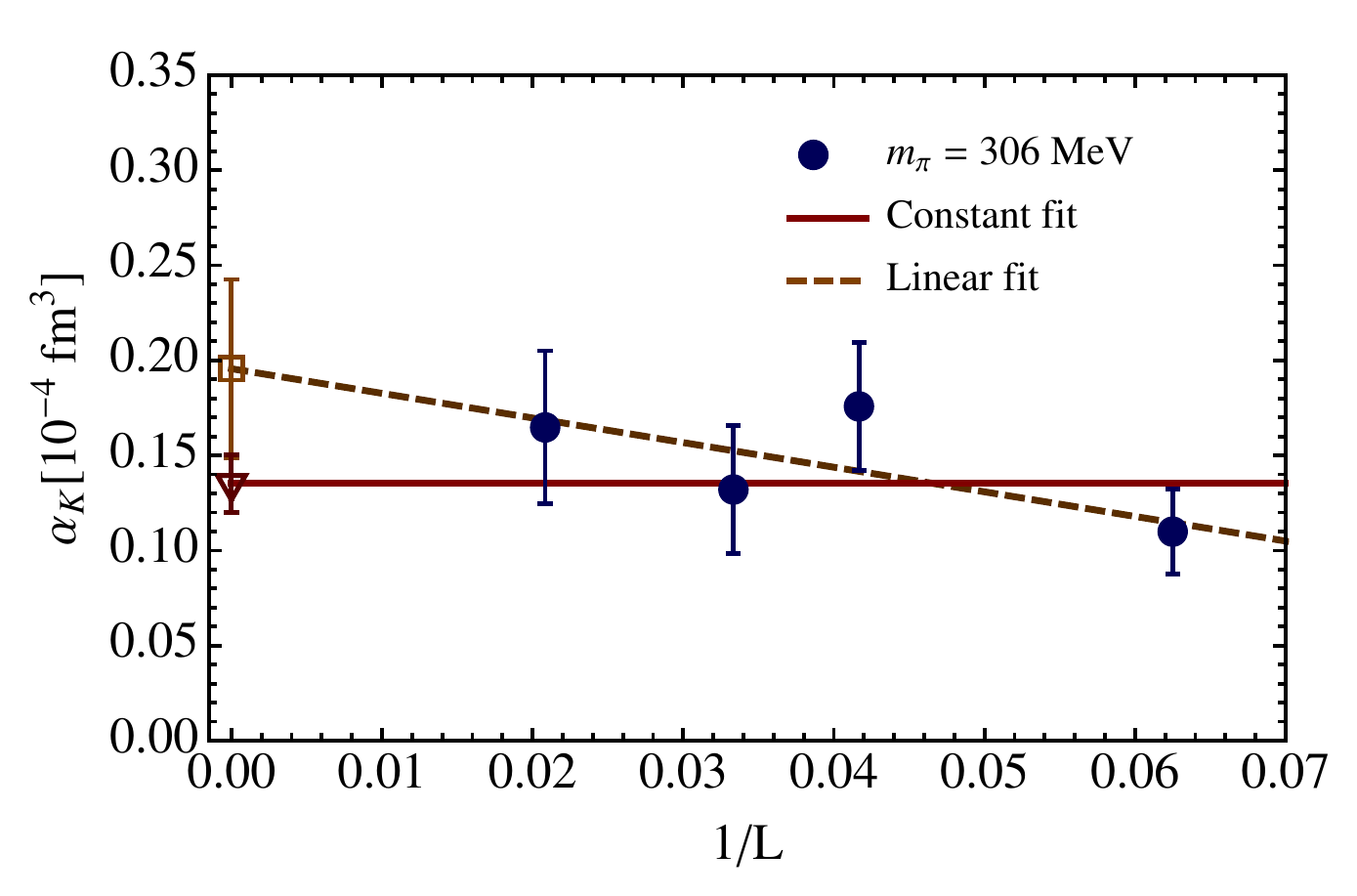}
\caption{Top: Neutral pion polarizability as a function of $1/L$. The right panel are our results for the 306 MeV ensemble and the left panel corresponds to the 227 MeV ensemble. On each plot we overlay our infinite volume extrapolations using a constant (red-solid line) and linear (orange-dash line) fit.
Bottom\,: Same as top graphs but for the neutral kaon.}
\label{pionkaonvol}
\end{figure}
%%%%%%%%

\begin{figure}[t]
 \centering
\includegraphics[width= 2.6 in]{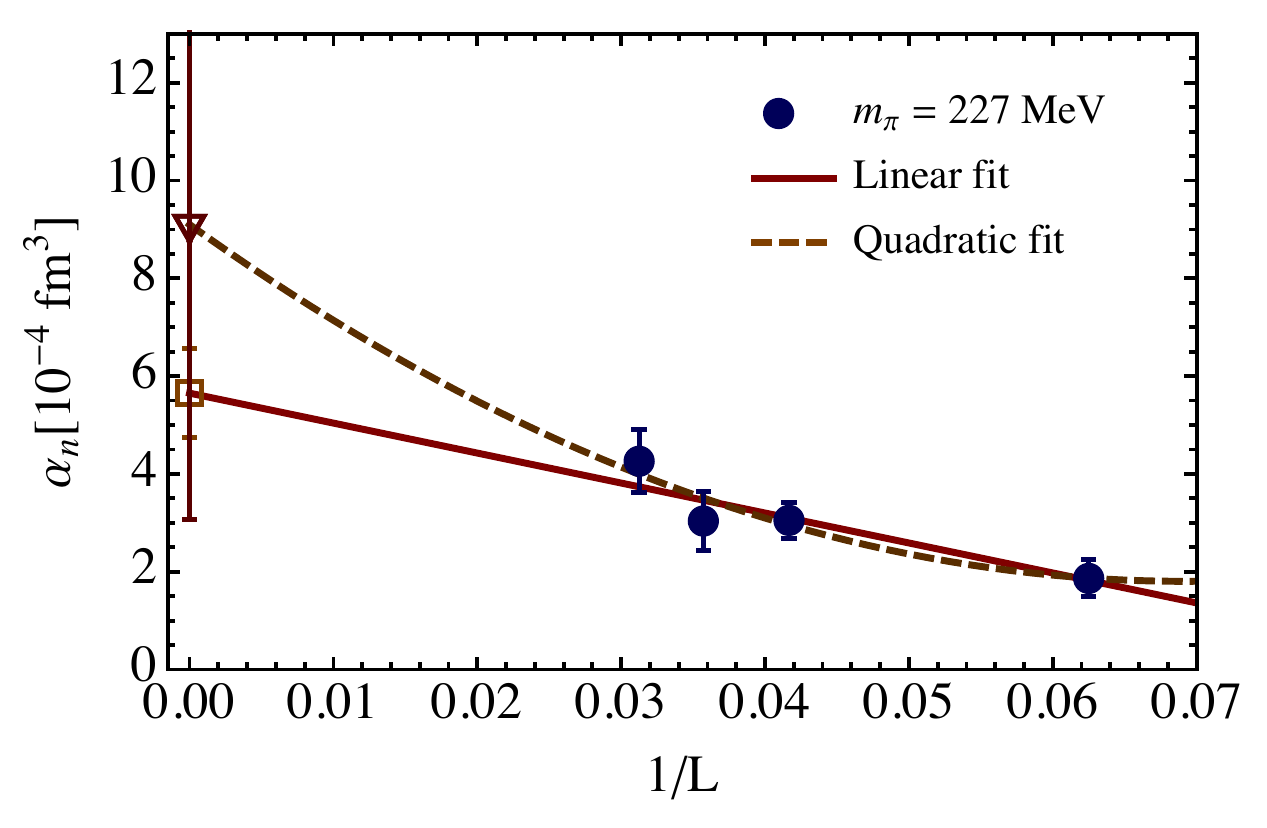}
\includegraphics[width= 2.5 in]{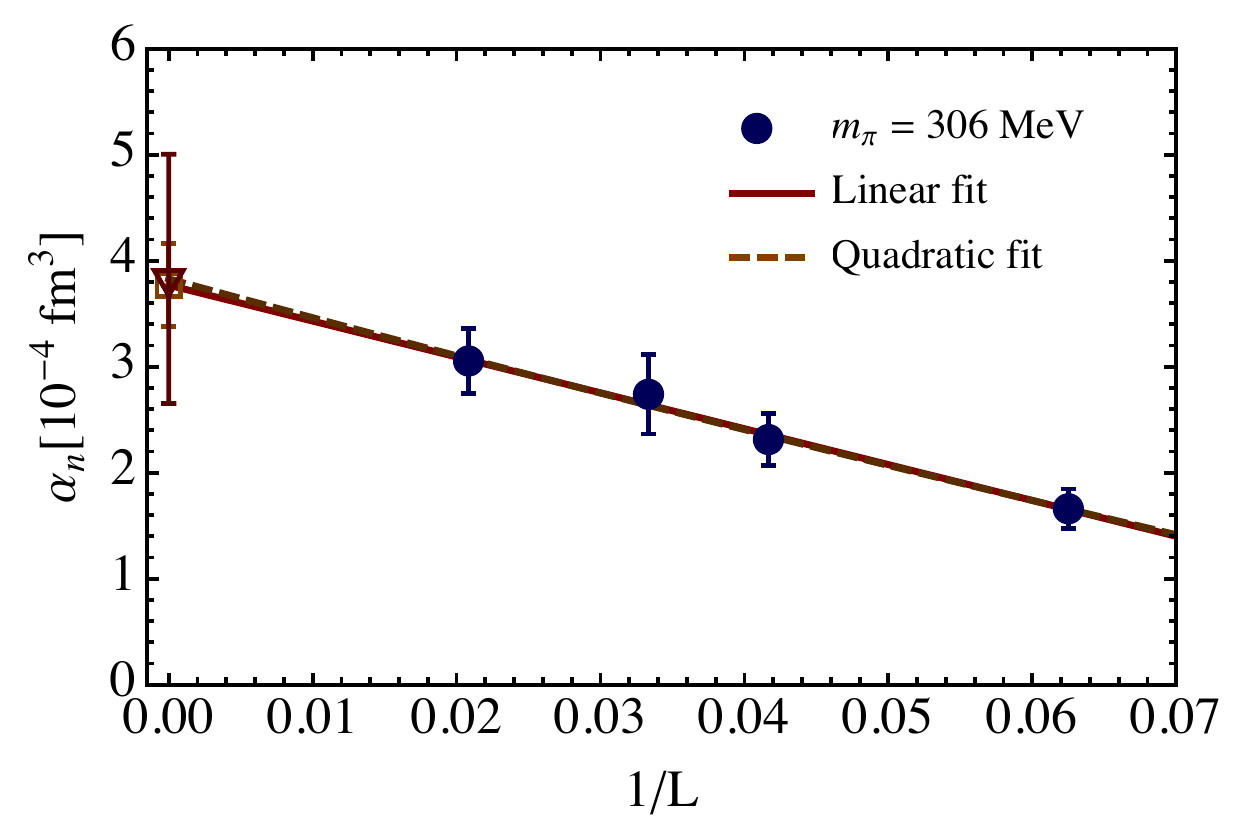}
\includegraphics[width= 2.6 in]{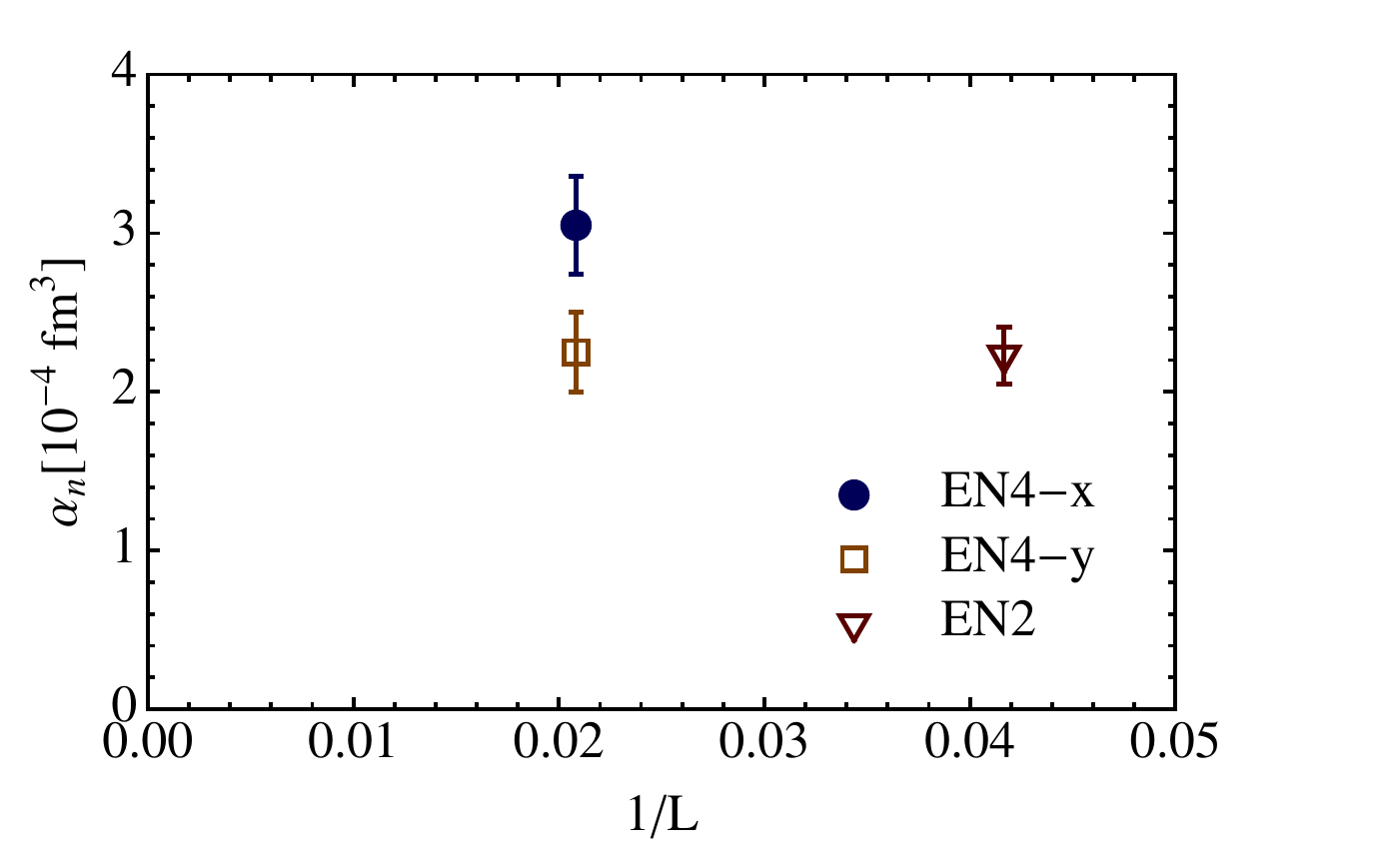}
\caption{Top: Infinite volume extrapolation for the neutron polarizability. Bottom:
Comparison of the $\EB$ ensemble to the results of the $\ED$ ensemble with the electric field in the $x$ direction and with the electric field in the  $y$ direction. See text for details.}
\label{neutronvol}
\end{figure}

The polarizabilities for the neutral pion, neutral kaon, and neutron were extracted for each ensemble using the fitting procedure described in Sec.~\ref{sec:fitprocedure}. For a given pion mass we studied the infinite volume behavior of the polarizability by extrapolating to the infinite volume limit using various polynomial degrees as fit models. We found, for the pion, that a constant extrapolation describes the data well. For the neutron and kaon we used a linear approximation.  Figures~\ref{pionkaonvol}~and~\ref{neutronvol} illustrate the results of the infinite volume extrapolation for both pion masses.

%%%%%%%%%
%%%%%%%%

The volume dependence analysis assumes that the finite size 
corrections are mainly driven by the extent of the lattice in the direction of the electric field
(which is in $x$-direction for this work). To verify this, we take our  $\ED$ lattice which has the 
spatial dimension $48\times24^2$ and place the electric field along the $y$-direction which has only 24 lattice units. 
We choose this ensemble because the difference in the $x$ and $y$ directions are the largest which gives us 
the best comparison. If the finite volume corrections associated with the transverse directions are small, we expect our results to be comparable to the results of the $\EB$~ensemble which has the  spatial 
dimension $24\times 24^2$. We display our results on the bottom panel of Fig.~\ref{neutronvol} for the $\EB$ lattice 
and the $\ED$ lattice for the electric field in the $x$ direction and the electric field in the $y$ direction. 
Our expectations are in very good agreement with our findings. That is, we found that dominant source of the finite 
size effects are connected to the extent of the lattice in the direction of the field.

%This result is statistically equivalent to the polarizability for the $\EB$ ensemble and statistically different from the case where we place the field along the $N_x=48$ direction.

%%%%%%%%%%%%%%%
%\begin{figure}[t]
%\centering
%\includegraphics[width= 2.6in]{neutron_compare_xy.eps}
%\caption{Comparison of the $\EB$ ensemble to the results of the $\ED$ ensemble with the electric field in the $x$ direction and with the electric field in the  $y$ direction. See text for details. }
%\label{plot:nvolplot2}
%\end{figure}
%%%%%%%%%%%%%%%

\subsection{Chiral Behavior}

\begin{figure}[t]
\centering
\includegraphics[width= 2.6in]{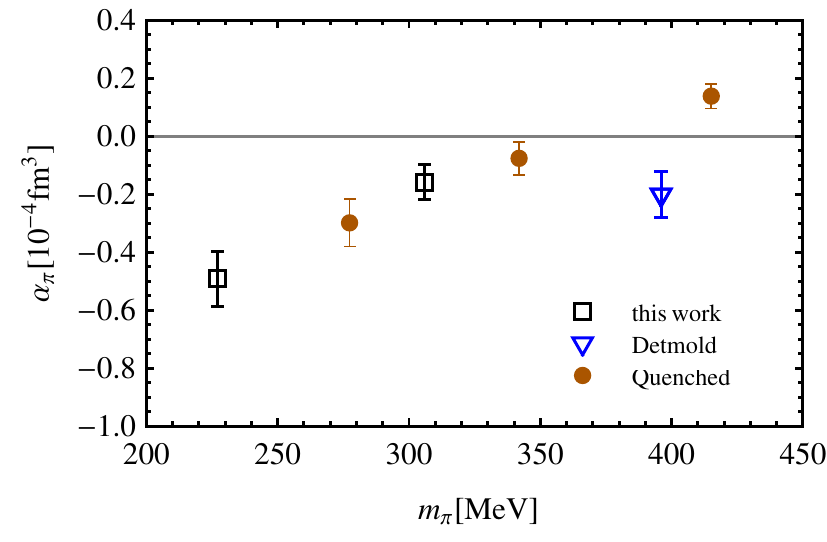}
\includegraphics[width= 2.5in]{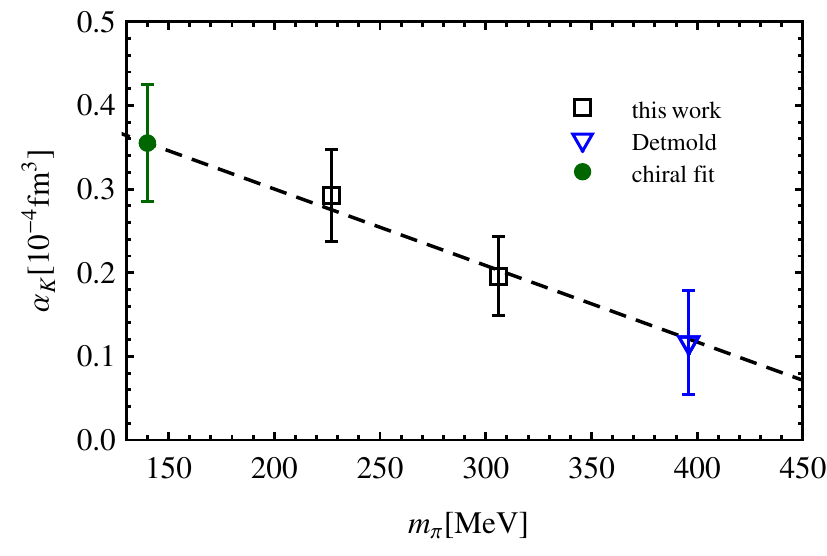}
\caption{Left panel: Pion mass dependence of the pion polarizability. The orange/circle points are quenched results found in~\cite{Alexandru:2010dx} and the blue/triangle point is the value determined in~\cite{Detmold:2009dx}. Right panel: Pion mass dependence for the kaon polarizability along with a chiral fit which includes the value determined in~\cite{Detmold:2009dx} }
\label{plot:chiralpionkaon}
\end{figure}

In this section we analyze the pion mass behavior for each of the particles we analyzed. On the left panel of  Fig.~\ref{plot:chiralpionkaon} we plot our infinite volume extrapolation results for the pion polarizability as a function of $m_{\pi}$. We also include the values calculated by Detmold {\it{et al.}}~\cite{Detmold:2009dx} at $m_{\pi} = 390$ MeV along with quenched results calculated by Alexandru and Lee~\cite{Alexandru:2010dx}.  The negative trend, which has been seen in our previous work~\cite{Lujan:2011ue}, is still present; determining its origin is an ongoing study. 

Our results for the kaon polarizability is illustrated on the right panel of Fig.~\ref{plot:chiralpionkaon}. 
The plot also includes the value determined by Detmold {\it{et al.}}~\cite{Detmold:2009dx}.
The kaon polarizability becomes larger as we lower the pion mass. In our previous work \cite{Lujan:2013qua} 
we performed an extrapolation to the physical point using only the $\EB$ and $\EEB$ lattices. We had found 
$\alpha_{\tiny{K}}= 0.269(43)\times10^{-4}$fm$^{3}$. Here we do the same analysis but now using our infinite 
volume results together with the value determined in~\cite{Detmold:2009dx}. 
We find $\alpha_{\tiny{K}}= 0.355(70)\times10^{-4}\, \text{fm}^3$. In this fit we assumed that the finite volume
corrections at $m_\pi=390$ MeV are small since they seem to decrease as we increase the pion mass.

%%%%%%%%
%%%%%

\begin{figure}[t]
\centering
\includegraphics[width= 2.7in]{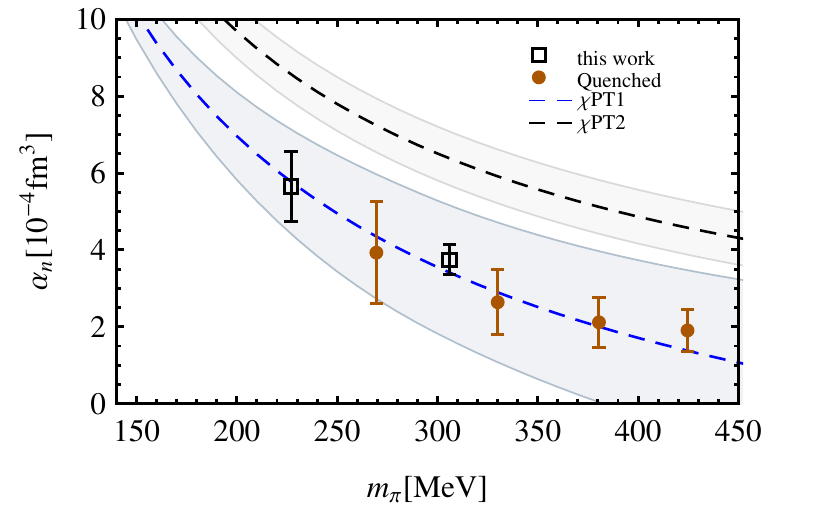}
\includegraphics[width= 2.7in]{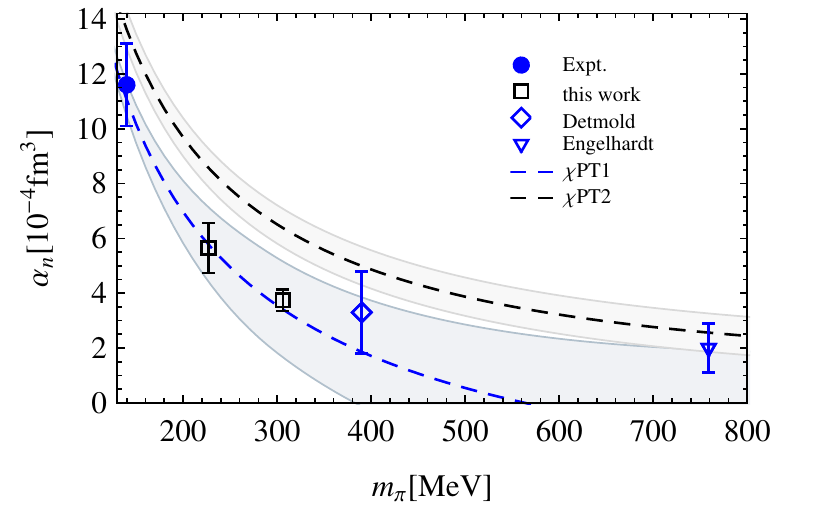}
\caption{Left panel: Pion mass dependence of the neutron polarizability. The orange/circle points are quenched results found in~\cite{Alexandru:2010dx}. The dashed lines are two different curves predicted by $\chi$PT$1$~\cite{Griesshammer:2012we,McGovern:2012ew} and $\chi$PT$2$~\cite{Lensky:2009uv}. Right panel: Plot of our results along with the experimental value and two other lattice calculations~\cite{Engelhardt:2007ub} and~\cite{Detmold:2010ts}. }
\label{plot:nvolplotchiral1}
\end{figure}

On the left panel of  Fig.~\ref{plot:nvolplotchiral1} we plot our infinite volume results for the neutron polarizability as a function of $m_{\pi}$ along with quenched data that were computed in~\cite{Alexandru:2010dx}.  We also compare our data to two different $\chi$PT predictions: $\chi$PT$1$~\cite{Griesshammer:2012we,McGovern:2012ew} and $\chi$PT$2$~\cite{Lensky:2009uv} to gauge systematic errors of our calculation. For a more detailed comparison of the two $\chi$PT predictions we refer the reader to~\cite{Lujan:2013qua}.

%These two $\chi$PT curves use different approximations in their calculations to derive the chiral form. In the case of $\chi$PT$1$ the calculation is expanded to N$^2$LO using a non-relativistic form for the propagators. There are two extra free parameters  which are determined by fitting to Compton scattering data. The second result, $\chi$PT$2$, includes terms up to NLO and uses relativistic propagators. They compute $\alpha_n$ as a function of $m_{\pi}$ with no free parameters. The error bars in $\chi$PT$1$ come from a careful analysis~\cite{Griesshammer:inprep} whereas the error bar for the second curve is fixed to a value estimated at the physical point. 

Similar to our findings in~\cite{Lujan:2013qua}, which only analyzed the $\EB$ and $\EEB$ ensembles, we find that our infinite volume results are compatible with the quenched ones. Moreover, our results are now in excellent agreement with the $\chi$PT$1$ curve.  This was not the case for our previous analysis which did not take into account the volume effects.  In Fig.~\ref{plot:nvolplotchiral1} we add the experimental point along with two other lattice calculations~\cite{Detmold:2010ts,Engelhardt:2007ub} for the neutron polarizability. Our results have the smallest pion masses used in polarizability studies and the smallest statistical errors. 

%%%%%
%%%%

%This analysis has shown that finite volume effects are very important for our calculations.
%For the pion masses used in our study, the calculation is nearly complete.  The systematics related to finite
%lattice spacing are expected to be small.  The only important systematic error comes from neglecting
%the charge of the sea quarks.  We can, to some degree, gauge this effect from work carried out by Detmold {\it{et al.}}~\cite{PhysRevD.73.114505} using $\chi$PT. In that work the authors derived a leading order $\chi$PT prediction for polarizability as a function of the sea quark masses.   They found that for pion masses between $140\,\text{MeV}$ and $300\,\text{MeV}$ the polarizability increases by $1.5$--$2\times 10^{-4}\,\fm^3$ when the charges of the sea quarks are set to their physical values.

% Figure~\ref{plot:nvplotcharge2} overlays our results with their $\chi$PT predictions with and without charged sea quarks. Note that in order to produce these curves the value of $|g_{N\Delta}| = 0.25$ was used;  this value is outside the expected range.  We used this value so that at the physical point the $\chi$PT prediction for $\alpha_n$ is $12 \times 10^{-4}\, \fm^3$. Our results, which were derived using neutral sea quarks, agree very well with this calculation. 

\section{Conclusion}\label{sec:conclusion}

We have presented a calculation of the electric polarizabilities for the neutron, neutral pion, and neutral kaon in the 
framework of lattice QCD.  We used two different pion masses (227 MeV and 306 MeV) to study the chiral behavior of 
the polarizability. Currently, these are the smallest masses used in polarizability studies. 
%The use of smaller pion 
%masses is important because we want to compare our results to $\chi$PT.  
We employed the background field method along with Dirichlet boundary conditions to place a constant electric 
field onto the lattice. 
A finite volume study was performed for each pion mass by computing the polarizability on four different lattice volumes. 
This was one of the most important results in this work. For the neutron we find that the finite volume corrections
are significant.
Without the infinite volume extrapolation our results did 
not agree with $\chi$PT calculations at our lowest pion mass where $\chi$PT is expected to be more accurate. 

\section{Acknowledgements} 
This work was done on the following GPU clusters: GWU IMPACT clusters, GWU CCAS Colonial One cluster, JLab clusters,   Fermilab clusters, and UK clusters.  This work is supported in part by the NSF CAREER grant PHY-1151648 and the U.S. Department of Energy grant DE-FG02-95ER-40907.

%%%%%%%%%%
\bibliographystyle{jhep-3}
\bibliography{my-references}

\end{document}